\documentclass[%
reprint,
superscriptaddress,
 amsmath,amssymb,
 showkeys,
aps,
prb,
floatfix,
]{revtex4-2}

\usepackage{threeparttable}
\usepackage{setspace}
\usepackage{makecell}
\usepackage{amsmath}
\usepackage{graphicx}
\usepackage{dcolumn}
\usepackage{bm}
\usepackage{hyperref}
\usepackage[mathlines]{lineno}
\usepackage{textcomp}
\usepackage{gensymb}
\usepackage{subfigure}
\usepackage{url}
\usepackage{xcolor}
\usepackage{hhline}
\usepackage[version=3]{mhchem}
\usepackage{multirow}

\begin{document}
\title{
Two-dimensional Ferroelectric \ce{Ga2O3} Bilayers with Unusual Strain-engineered Interlayer Interactions
}

\author{Junlei Zhao}
\email{zhaojl@sustech.edu.cn}
\affiliation{Department of Electrical and Electronic Engineering, Southern University of Science and Technology, Shenzhen 518055, China}

\author{Xinyu Wang}
\affiliation{Department of Electrical and Electronic Engineering, Southern University of Science and Technology, Shenzhen 518055, China}

\author{Haohao Chen}
\affiliation{Department of Electrical and Electronic Engineering, Southern University of Science and Technology, Shenzhen 518055, China}

\author{Zhaofu Zhang}
\email{zz389@cam.ac.uk}
\affiliation{Department of Engineering, University of Cambridge, Cambridge CB2 1PZ, United Kingdom}

\author{Mengyuan Hua}
\email{huamy@sustech.edu.cn}
\affiliation{Department of Electrical and Electronic Engineering, Southern University of Science and Technology, Shenzhen 518055, China}


\begin{abstract}

Two-dimensional (2D) van der Waals (vdW) materials and their bilayers have stimulated enormous interests in fundamental researches and technological applications. 
Recently, a group of 2D vdW \ce{III2-VI3} materials with out-of-plane ferroelectricity have attracted substantial attentions. 
In this work, the structural, electronic and optical properties of 2D ferroelectric \ce{Ga2O3} bilayer system are systematically studied using \textit{ab-initio} computational method. 
Intrinsic dipoles of the two free-standing monolayers lead to three distinct dipole models (one ferroelectric and two antiferroelectric models). 
The stable stacking configurations of ferroelectric and antiferroelectric dipole models can be transferred with polarization reversal transition of the monolayers without additional operation. 
Interlayer perturbation effects combined with biaxial-strain engineering lead to high tunablility of the electronic and optical properties of the bilayer systems.   
Surprisingly, the results reveal a phase transition from vdW to ionic interlayer interaction induced by in-plane biaxial tensile strain.
Detailed analyses suggest a transition mechanism based on the ionic bonding nature of the \ce{Ga2O3} system, involving interlayer rearrangement of anions to compensate the symmetry breaking of the heavily distorted ionic folding configurations.
These insights can open new prospects for future experimental synthesis, characterization and application of 2D \ce{Ga2O3} atomic-thin layered systems.

\end{abstract}

\maketitle

\section{Introduction} \label{sec:intro}

Ten years after the first synthesis of graphene~\cite{novoselov2004electric}, Geim and Grigorieva highlighted a promising strategy of using two-dimensional (2D) van der Waals (vdW) materials as the ``atomic-scale Lego blocks" to build precisely controlled lamellar structures with designed optical and electrical characteristics~\cite{geim2013van}.
To fulfill this prospect, tremendous efforts in the past decade have been put in exploring of different 2D vdW materials and their multilayer structures, resulting an incredibly diverse library of several major 2D vdW families, such as hexagonal boron nitride ($h$-\ce{BN})~\cite{park2014hbn}, transition metal dichalcogenides (TMDCs)~\cite{mak2010atomically, wang2012electronics}, MXenes~\cite{naguib2014mxene}, other single-element crystals (\textit{e.g.}, silciene~\cite{laimi2010silicene} and phosphorene~\cite{liu2014phosphorene}), and layered metal oxides~\cite{zavabeti2017liquid, zhang2021hexagonal}.
By vertically stacking two or more 2D vdW ``Lego blocks" together, emergent applications can be realized in the fields of electrodes~\cite{ko2021robust}, transistors~\cite{chen2021logic}, optical sensors~\cite{hong2021highly}, and surface catalysis~\cite{deng2016catalysis, guo2020theoretical}.
More fundamentally, 2D vdW multilayer structures serve as vital and active frontiers for several recent exciting advances in condensed matter physics, such as unconventional superconductivity~\cite{cao2018unconventional}, topological insulator~\cite{fatemi2018electrically}, searching of Majorana zero modes~\cite{jack2021detecting}, ultralow interlayer friction~\cite{shen2021ultralow}, interlayer excitons~\cite{ciarrocchi2019polar, bi2021excitonic}, and valleytronics~\cite{xiao2012coupled, schaibley2016valleytronics}.

Amongst different 2D vdW materials, 2D metal oxides can be synthesized in metastable stoichiometric structures which differ from their conventional bulk counterparts~\cite{zhang2021hexagonal}.
Typical 2D metal oxides, such as \ce{ZnO}~\cite{tusche2007observation, zhang2020strain},  \ce{TiO2}~\cite{tao2011a, zhang2021hexagonal}, and $\gamma$/$\beta$-\ce{Ga2O3}~\cite{yang2019hydrothermal, li2021template}, are often adopted to centrosymmetric structure without polarization, owing to the well-known ``polar instability" effect~\cite{noguera2000polar}.
However, ferroelectrics can be achieved in some 2D monolayers~\cite{fei2018ferroelectric, yang2018origin, xiao2018intrinsic}.
Such 2D monolayers consist of charged planes of cations and anions, which can generate a spontaneous electric field like a capacitor.
By switching the polarization, distinct ``ON" and ``OFF" states can be controlled precisely, which are highly desirable in applications such as surface catalysis~\cite{kim2021computational}, gas sensing~\cite{tang2020reversible} and novel electronics~\cite{wu2020high, ding2021antiferro}.

Our previous studies~\cite{liao2020tunable, SFzhao2021two, SFzhao2021phase} revealed a metastable configuration of 2D vdW \ce{Ga2O3} monolayer with intrinsic dipole pointing from top to bottom surfaces (\textit{i.e.}, from negatively to positively charged centers).
The structure is analogous to ferroelectric zinc blende (FE-ZB$'$) phase of \ce{III2-VI3} compounds~\cite{ding2017prediction, xiao2018intrinsic}, as shown in Supporting Information Figure S1.
The \ce{Ga2O3} monolayer is stacked in the order of O-Ga-O-Ga-O, forming a quintuple-atomic-layer and hexagonal-symmetry structure.
Although FE-ZB$'$ phase is metastable (0.043 eV per atom higher than monolayer $\beta$-\ce{Ga2O3} phase (see Ref. \citenum{SFzhao2021phase}), it is highly likely to be observed under substrate confinement, external electric field or kinetic trapping effect of the synthesis methods.
Therefore, potential 2D vdW bilayer structure based on FE-ZB$'$ \ce{Ga2O3} monolayer is of great interest and can provide deeper insights in the 2D ferroelectric systems.

In this work, we use first-principles calculation to study the 2D vdW FE-ZB$'$ \ce{Ga2O3} bilayer systems.
In the first part, depending on the directions of the intrinsic dipoles, one ferroelectric and two antiferroelectric models are constructed and the most stable stacking layouts are systematically investigated.
Intriguingly, although bulk \ce{Ga2O3} is a non-vdW material with high ionicity, the FE-ZB$'$ bilayer systems exhibit typical vdW interactions at their relaxed states.
In the second part, we study the electronic and optical properties of the three stable bilayer structures and further discuss on the effects of interlayer perturbation and strain engineering. 
In the last part, an unexpected phase transition from vdW to ionic interlayer interaction, induced by in-plane biaxial tensile strain, are revealed.
We closely study the phase transition mechanism with detailed analyses of binding energies, interlayer distances, charge density distribution and ionic coordination.

\section{Methods -- Computational Details} \label{sec:methods}

The \textit{ab-initio} calculations were conducted based on density functional theory (DFT) implemented with Vienna Ab initio Simulation Package (VASP)~\cite{vasp1993, vasp1996}.
A high planewave cutoff energy of 700 eV was used to obtain well converged results.
Energy convergence criterion was 1$\times$10$^{-7}$ eV for electronic self-consistent field iteration and the force convergence criterion was 1$\times$10$^{-3}$ eV/\r A for ionic structural optimization.
The first Brillouin zone was sampled with the dense $\Gamma$-centered reciprocal grid of 13$\times$13$\times$1.
Gaussian approximated smearing with a width of 0.03 eV was used to determine the partial occupancies.
Perdew-Burke-Ernzerhof (PBE) version of the generalized gradient approximation exchange-correlation functional~\cite{pbe1996, pbe1997} was used for studying structural properties and biaxial strain induced interlayer transition, while the Heyd-Scuseria-Ernzerhof (HSE06) hybrid functional~\cite{hse2006ori, hse2006solid} with standard 25\% of the exact Hartree-Fock exchange\cite{hf1928a, hf1930} was used when calculating electronic and optical properties. 
The vdW correction with Grimme’s scheme (DFT-D3)~\cite{dftvdw2010} was considered.
A linear dipole correction is added to the local potential to correct the errors caused by periodic boundary condition in $z$ axis.
Some of the post-processing tasks were done with VASPKIT package~\cite{vaspkit2021}.
OVITO~\cite{ovito2010} and VESTA~\cite{vesta2011} packages were used for visualizing the atomic configurations and charge densities. 

Eighteen ten-atom hexagonal unit cells of the three dipole models (six stacking configurations for each model) were constructed with the initial interlayer distance of 3 \r A. 
The in-plane lattice constants and the internal positions of the nuclei were optimized by using conjugate gradient method. 
To avoid interaction between the periodic images in the $z$ direction, a vacuum layer of 25 \r A was appended to the bilayer systems. 
The lattice constants of the bilayers optimized by the PBE functional are $a=b\simeq3.070$ \r A for all the calculated configurations with only marginal differences in the third decimal values. The detailed formulas of the post-calculation analyses can be found in Supporting Information. 

\section{Results and Discussion} \label{sec:results}

\subsection{Structural properties} \label{sec:struc}
  
Before systematically studying the structural properties of the FE-ZB$'$ \ce{Ga2O3} bilayer systems, we first review the structure of the monolayer as shown in Supporting Information Figure S1. 
For clarification, three atomic layer stacking sites $A$, $B$, and $C$ are defined in the \ce{Ga2O3} monolayer, where (i) the site $A$ aligns to the top hollow ($A_{\ce{H}}$) position, the center \ce{O} ($A_{\ce{O}}$) and bottom \ce{Ga} ($A_{\ce{Ga}}$) layers; (ii) the site $B$ to the top \ce{Ga} ($B_{\ce{Ga}}$) layer and the bottom hollow ($B_{\ce{H}}$) position; and (iii) the site $C$ to the top and bottom \ce{O} ($C_{\ce{O}}$) layers. 
The FE-ZB$'$ \ce{Ga2O3} monolayer has a special polarization reversal transition where the central layer of \ce{O} atoms jump from site $A$ to site $B$ with the transition barriers of 0.44 eV~\cite{SFzhao2021phase}, as illustrated in Supporting Information Figure S1. 

In the FE-ZB$'$ \ce{Ga2O3} bilayer systems, the orientations of the upper and lower monolayers are defined by the directions of the out-of-plane dipole moments (within a monolayer, pointing from top to bottom surfaces, see Supporting Information Figure S1).
Therefore, the free-standing \ce{Ga2O3} bilayers have three dipole models, \textit{i.e.}, Up--Up (U--U), Down--Up (D--U) and Up--Down (U--D) models. 
The Down--Down (D--D) model is physically identical to the U--U model, so it will be discussed as the U--U model herein. 
The U--U model is ferroelectric (FE), whereas the D--U and U--D models are anti-ferroelectric (AFE) with the dipole moments pointing inwards and outwards, respectively,  thus the overall polarization is canceled out.

\begin{figure*}[ht!]
 \centering
 \includegraphics[width=17.2cm]{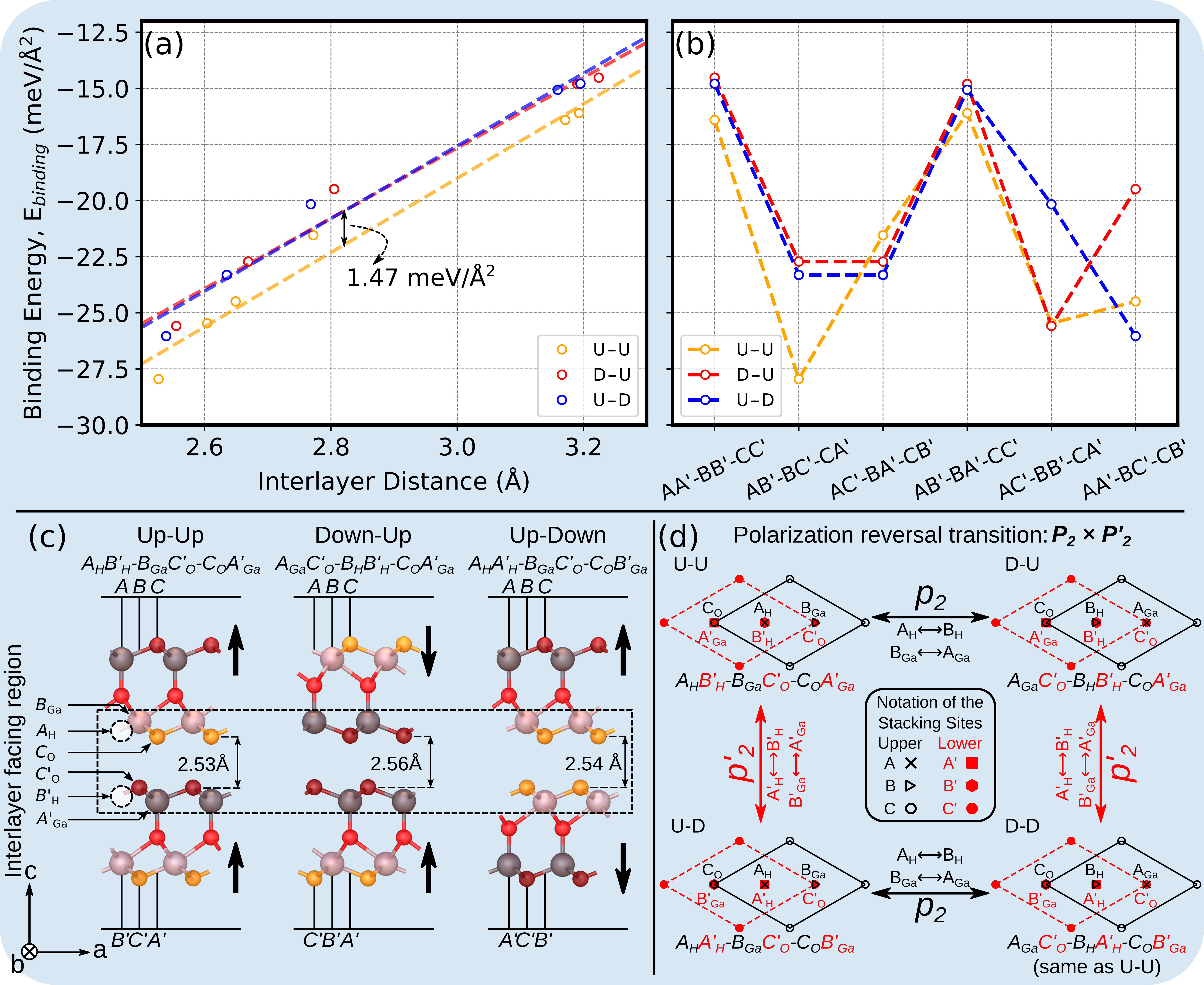}
 \caption{Structure properties: (a) An approximately linear correlations are seen between the interlayer distances and the binding energies for the three models. (b) A plot against the stacking index. (c) Side views of the atomic structures of the lowest-energy \ce{Ga2O3} bilayer configurations. (d) Polarization reversal transitions among different dipole models. Black and red symbols represent the upper and lower monolayers, respectively. See Supporting Information Figure S3 for the detailed group analysis.}
 \label{fig:struc}
\end{figure*}

We construct all six possible high-symmetric stacking configurations for the three dipole models.
The stacking configurations are obtained by rotational ($R$) and/or translational ($\mathbf{t}$) operations of the lower monolayer with respect to the upper monolayer.
In this way, the six stacking configurations can be indexed by the vertical alignment of the stacking sites between the upper ($A$, $B$ and $C$) and lower ($A'$, $B'$ and $C'$) monolayers. 
The binding energies, $E_\mathrm{Binding}$ and the interlayer distances (defined as the vertical distance between two interlayer $C_{\ce{O}}$ atoms, see Figure~\ref{fig:struc}c) of all the eighteen structures are summarized in Supporting Information Table S1. 
All calculated configurations can be found in Supporting Information Figure S2.
The binding energies ranging from -14.524 to -27.952 meV/\r A$^{2}$ are consistent with the previous high-accuracy calculations on the weakly bonded layered compounds using different vdW-corrected functionals~\cite{bjorkman2012van, mounet2018two}, indicating a typical vdW interlayer interaction. 

A close analysis on the data reveals two intriguing facts: 
(i) An approximately linear correlation is seen between the binding energies and interlayer distances for the three models, as shown in Figure~\ref{fig:struc}a, and the $E_\mathrm{binding}$ of the U--U model is 1.47 meV/{\r A}$^{2}$ larger than the ones of the D--U and U--D models at the same layer distance; 
(ii) The plot against the stacking indices reveal two energy degenerated configurations ($AB'-BC'-CA'$ and $AC'-BA'-CB'$) for the D--U and U--D models, however, it is not seen in the U--U model, as shown in Figure~\ref{fig:struc}b. 

To elucidate the physical mechanism behind these facts, we closely examine the stacking configurations with respect to the interlayer facing region. 
As shown in Figure~\ref{fig:struc}c, the lowest-energy stacking configurations are labeled with the interlayer facing atoms indicated by the dashed boxes. 
In the U--U model, the facing sites are $A_{\ce{H}}$/$B'_{\ce{H}}$, $B_{\ce{Ga}}$/$C'_{\ce{O}}$ and $C_{\ce{O}}$/$A'_{\ce{Ga}}$.
The interlayer facing sites of the D--U and U--D models can be defined similarly. Clearly, it is energetically favorable to have two cation/anion (\ce{Ga}/\ce{O}) and one hollow/hollow (\ce{H}/\ce{H}) facing pairs for all the three models. All the $C_{\ce{O}}/C'_{\ce{O}}$ stacking configurations are not stable since the anions get relatively close to each other. The zigzagged facing surfaces (with $C_{\ce{O}}$ atoms tilt-out) further lead to the fact (i) that the most stable configurations have the shortest interlayer distances, while the most unstable ones with the $C_{\ce{O}}/C'_{\ce{O}}$ facing pair have the longest distances. 
On the other hand, it also provides useful insights to the explanation of fact (ii). 
Because the two energy degenerated configurations of the D--U and U--D models have exactly the same facing pairs, \textit{i.e.}, \ce{Ga}/\ce{H}, \ce{H}/\ce{O}, and \ce{O}/\ce{Ga}. In fact, they are centrosymmetric with respect to the center of the interface, however, obviously, this centrosymmetry is broken in the FE U--U model (see Supporting Information Figure S2).      

In experimental conditions, controlling the stacking configuration of a 2D vdW bilayer system relies on physical transformations such as gliding, rotating and flipping operations on a single monolayer. 
However, specifically for the FE-ZB$'$ \ce{Ga2O3} bilayer system, the aforementioned polarization reversal transition can result in a change of the stacking configurations as well. 
In Figure~\ref{fig:struc}d, we show a symbolic top-view map of such transitions with labeled interlayer facing sites. In this way, the transition can be written in the notations of the switched stacking sites as ($A_{\ce{H}} \longleftrightarrow B_{\ce{H}}$)/($B_{\ce{Ga}} \longleftrightarrow A_{\ce{Ga}}$).
Interestingly, the polarization reversal transition of a monolayer will bring the bilayer system from one stable stacking configuration to another stable configuration without additional translation. 
For example, starting from the lowest-energy $A_{\ce{H}}B'_{\ce{H}}-B_{\ce{Ga}}C'_{\ce{O}}-C_{\ce{O}}A'_{\ce{Ga}}$ configuration of the U--U model, the transitions of the upper monolayer will lead to the lowest-energy configuration of the D--U model, \textit{vice versa}. The same case can be seen in the transition from the U--U to U--D models.    
Therefore, we further make a comprehensive symmetry group table with 24 objects as shown in Supporting Information Figure S3, which consists of a Dihedral group ($\mathbf{D}_{3}$) multiplying two reversal transition groups ($\mathbf{P}_{2} \times \mathbf{P'}_{2}$).   
The group analysis reveals that the polarization reversal transition of a monolayer is equivalent to a flipping operation followed by a 60$\degree$-rotation on this monolayer. 
We note that this symmetry group, in principle, can be generalized to other FE-ZB$'$ \ce{III2-VI3} bilayer systems such as \ce{In2Se3}~\cite{ding2017prediction, li2020band, yang2021an}. Therefore, utilizing the polarization reversal transition to control the interlayer stacking configurations can be a promising strategy for future experimental studies.    

After investigating the structural properties of the FE-ZB$'$ \ce{Ga2O3} bilayer systems, we further select the lowest-energy configurations of the three dipole models and study their electronic, optical and strain-engineered properties in the following sections. For concision, herein we will use the U--U, D--U and U--D models to refer to their stable stacking configurations, if without any other specification. 

\subsection{Electronic properties} \label{sec:electronic}

\begin{figure*}[ht!]
 \centering
 \includegraphics[width=17.2cm]{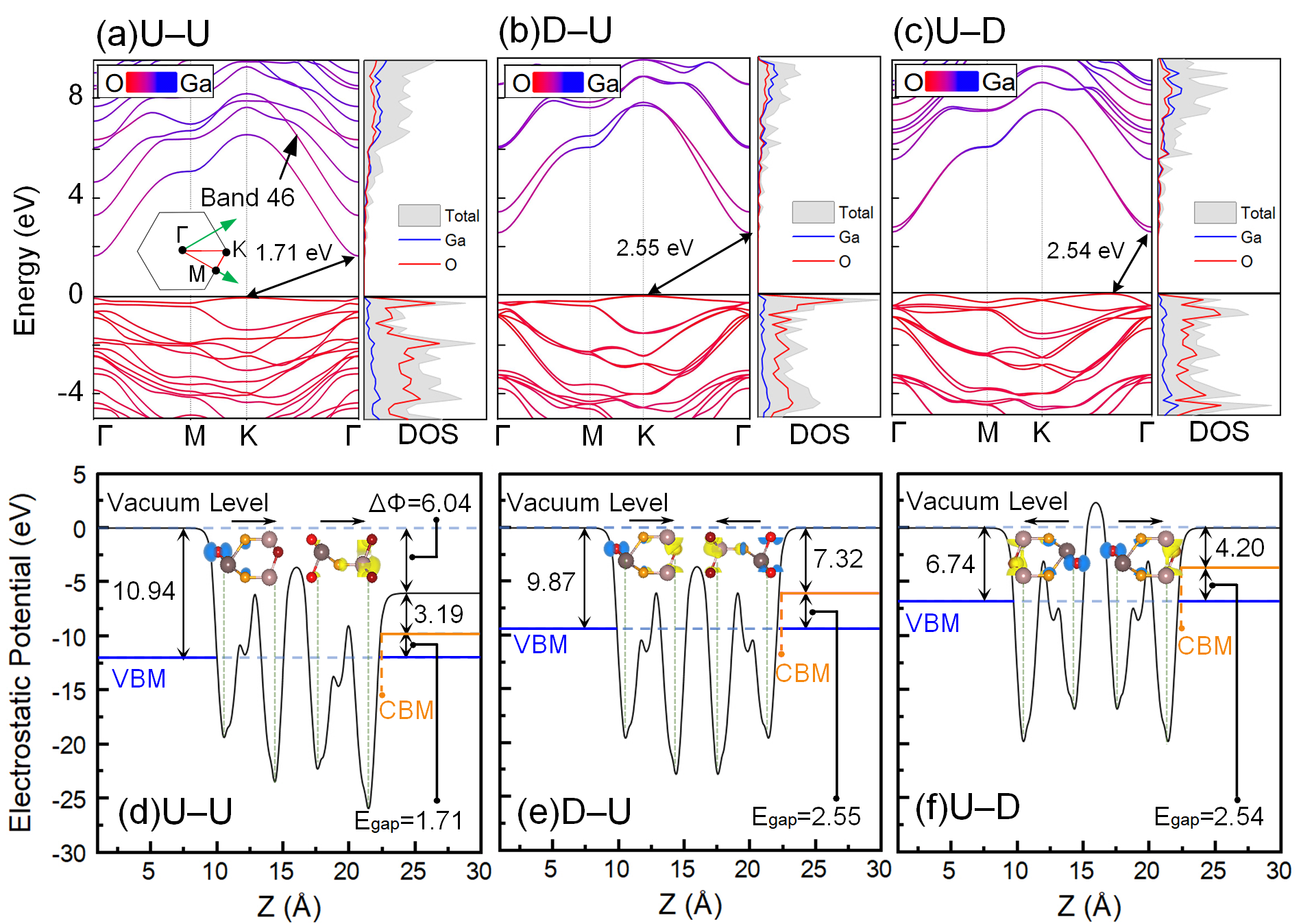}
 \caption{(a-c) The band structures (left panels) and DOS (right panels) of the three stable dipole models calculated by HSE06 hybrid functional. The CBMs are leveled to zero eV. In the left panels, red and blue color coding represents the contribution strength from \ce{O} and \ce{Ga}, respectively. The black arrows indicate the indirect band gaps from VBMs to CBMs. Band 46 indicates the shifted CBM of the lower monolayer in the U--U model. (d-f) The planar-average electrostatic potential curves along the $z$ axis. The inset atomic structures are aligned to the local minima of the potential by the dashed green lines, and the spatial distribution (squared wave function) of the CBMs (yellow) and VBMs (blue) are visualized. Different vacuum levels (dashed light blue), relative levels of VBMs (solid blue) and CBMs (solid orange) are shown.}
 \label{fig:pot}
\end{figure*}

The HSE06 band structures of the three stable stacking configurations (left panels) 
are shown in Figure~\ref{fig:pot}a-c, together with the corresponding total/partial density of state (DOS, right panels). 
Indirect band gaps from valence band maximum (VBM) to conduction band minimum (CBM) of 1.71, 2.55 and 2.54 eV are seen for U--U, D--U and U--D models, respectively.
Compared with the band gap ($\sim$2.85 eV with the lattice constants of 3.070 \r A) of the \ce{Ga2O3} monolayer~\cite{liao2020tunable}, we note that the band gaps of the bilayer systems is overall smaller. 
The VBMs of the U--U and D--U models are located at the high-symmetric $K$ point, while the VBM of the U--D model lies in between the $K$ and $\Gamma$ points.
The CBMs are located at $\Gamma$ point for all three models. 
The element-weighted color coding reveals that the VB is contributed mainly by oxygen $2p$ orbitals, while the CB by gallium $4s$ orbitals (see Supporting Information Figure S5). 

In general, electronic band structure of a bilayer system can be considered as the overlapping of two individual monolayers with the first-order approximated perturbations caused by the interlayer synergetic effects~\cite{lopes2007graphene}.
Charge redistribution in the interlayer facing region (and even beyond that) can lead to splitting, shifting and deforming of the band edges~\cite{novoselov2016heterostructures}. 
In the FE-ZB$'$ \ce{Ga2O3} bilayer systems, these interlayer effects are significantly enhanced because of the intrinsic dipole and the unique spatial locations of the VBM and CBM.  

As shown in Figure~\ref{fig:pot}d-f, the planar-average electrostatic potentials and the spatial distributions of the CBMs and VBMs can provide deeper insights on the understanding of such interlayer effects, together with the plots of the band structures. 
For the FE U--U model, the intrinsic dipole of the lower monolayer acts as an external field to shift the energy bands of the upper monolayer down about 1.67 eV (the energy difference between the ``Band 46" and the CBM at the $\Gamma$ point). 
Therefore, energetically, the original VB edge of the lower monolayer and the CB edge (``Band 46" in Figure~\ref{fig:pot}a) of the upper monolayer are now embedded in the VB and CB of the bilayer system. 
In addition, as shown in Figure~\ref{fig:pot}d, these embedded band edges of the two monolayers are spatially located at the interlayer facing region, whereas the VBM and CBM of the bilayer are now located at the two opposite outward surfaces.   
Moreover, because the outward surfaces are not affected by the interlayer effects, the shapes of the VB and CB edges are almost identical to the ones of the monolayer as shown Supporting Information Figure S6a. 

In contrast, for the AFE D--U and U--D models, it is clear that the shifting effect owning to the intrinsic dipoles are canceled by each other, which leads to a similar band gap comparing to the one of the monolayer. 
However, the interlayer charge redistribution results in an energy split of the CB edge of the D--U model, and a deformation of the VB edge of the U--D model. 
These effects can be explicitly confirmed by a close comparison of the band structures as shown in Supporting Information Figure S6b,c. 

To explore the electrical properties of the \ce{Ga2O3} bilayer systems, we evaluate the carrier transport capacity by calculating the carrier effective mass and mobility. 
The effective masses of the electron and hole can be calculated by fitting the band curves near the CBM and VBM, respectively. 
The hole effective mass of the \ce{Ga2O3} bilayer is more than 20 times larger than the electron effective mass. 
This is very similar to the properties of the bulk $\beta$-phase \ce{Ga2O3} as well as its monolayer systems~\cite{peelaers2015brillouin, peelaers2017lack}.  
Therefore, only the effective mass and mobility of electron are analyzed, as summarized in Table~\ref{tbl:notes}.
 
\begin{table*} 
  \caption{Electron mobilities, $\mu$, of the \ce{Ga2O3} bilayer systems along the $x$ and $y$ directions. $m_{0}$ is the free electron mass. The directions of $x$ and $y$ are shown in Supporting Information Figure S7c.}
  \label{tbl:notes} 
  \begin{tabular}{ m{1.5cm} | m{1.5cm} | m{1.5cm} | m{1.5cm} | m{1.5cm} | m{1.5cm} | m{2cm} | m{2cm} }
    \hline
    \hline
    Dipole model & $m_{e}^{*}$ ($m_{0}$) & $E_{dpc}^{x}$ (eV) & $E_{dpc}^{y}$ (eV) & $C_\mathrm{2D}^{x}$ (J/m$^{2}$) & $C_\mathrm{2D}^{y}$ (J/m$^{2}$) & $\mu_{x}$ (cm$^{2}$/V$\cdot$s) & $\mu_{y}$ (cm$^{2}$/V$\cdot$s) \\
    \hline
     U--U & 0.303 & -2.451 & -2.532 & 241.946 & 221.585 & 9357 & 8028 \\
     D--U & 0.306 & -3.293 & -2.989 & 219.440 & 201.426 & 4610 & 5137 \\
     U--D & 0.286 & -3.264 & -2.967 & 232.685 & 232.663 & 5696 & 6895 \\
    \hline
    \hline
  \end{tabular}
\end{table*}

The electron effective mass, $m_{e}^{*}$, of the bilayer system is isotropic as expected, similar to our previous calculation of the monolayer \ce{Ga2O3} system and other members of the 2D vdW \ce{III2-VI3} family~\cite{fu2018intrinsic}. 
The isotropy can also be seen from the same curvatures of the CB edges near $\Gamma$ point along $\Gamma$-$K$ and $\Gamma$-$M$ paths. 
The electron mobilities of the bilayer systems are quite comparable to the ones of the \ce{Ga2O3} monolayer ($\sim$6000 cm$^{2}$/V$\cdot$s)~\cite{liao2020tunable}. 

\subsection{Strain-engineered electronic and optical properties} \label{sec:strain}

\begin{figure*}[ht!]
 \centering
 \includegraphics[width=14cm]{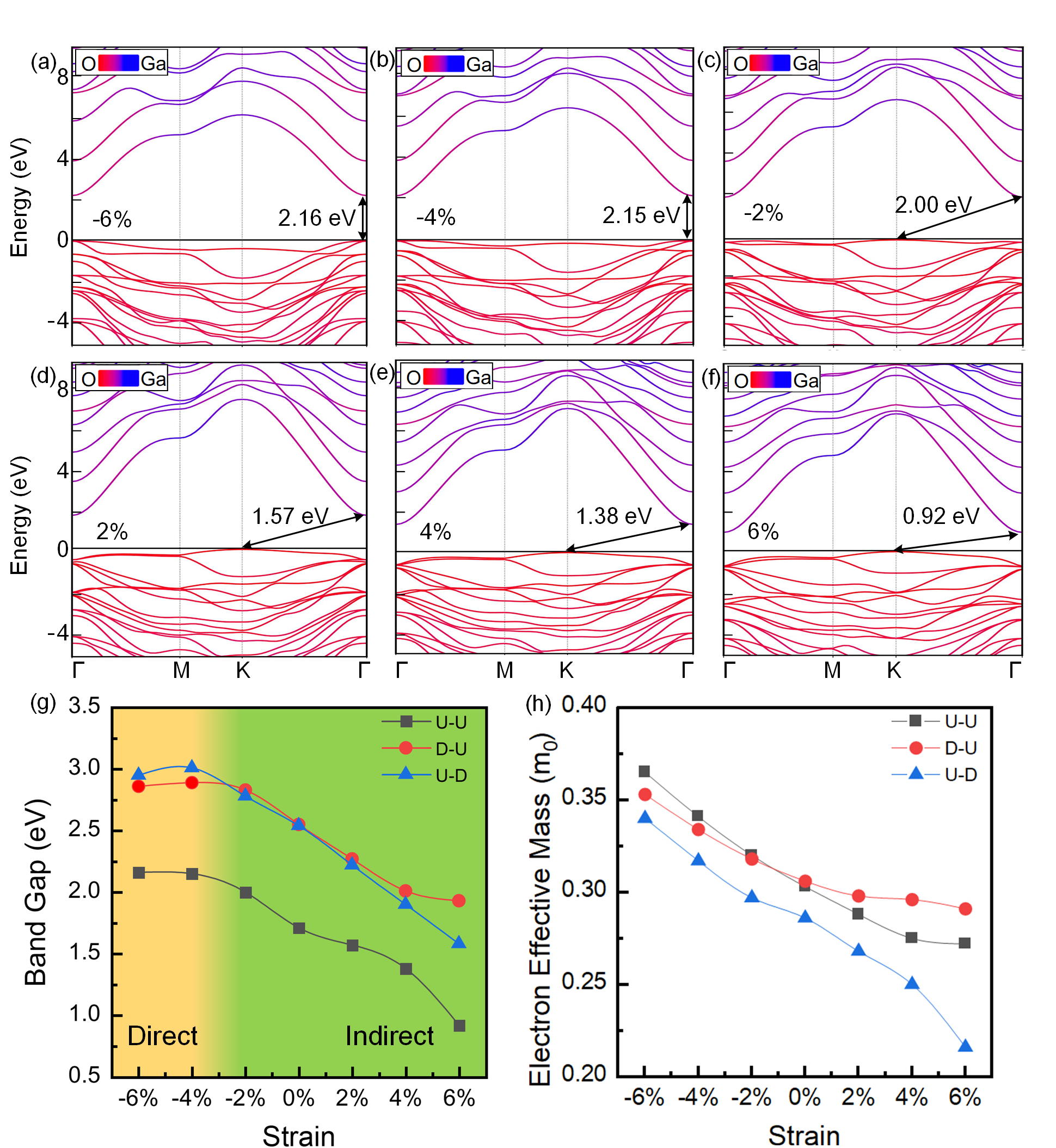}
 \caption{(a-f) Band structures of the U--U model with biaxial strains from -6 to 6\%. (g,h) Biaxial strain induced changes in band gaps, and electron effective masses of the three models. The yellow-to-green shadow in (g) indicates a direct-to-indirect transition of the band gaps between the -4 and -2\% strains.}
 \label{fig:u-u}
\end{figure*}

An attractive application of 2D vdW materials is flexible devices with strain-engineered electronic and optical properties~\cite{daus2021high, ko2021robust}, so we further explore the controllability of the bilayer systems via biaxial strain engineering. 
Figure~\ref{fig:u-u}a-f show the change of the band structure of the U--U model upon the biaxial strain from -6 to 6\%. The corresponding band structures of the D--U and U--D models are shown in Supporting Information Figure S8,S9. 
A direct-to-indirect transition of the band gaps are observed for the three models between the -4 and -2\% strains, as indicated in Figure~\ref{fig:u-u}g. 
A close inspection on the band structures reveals a similar transition for the three models, that is, the CBMs remain unchanged at the $\Gamma$ point, while the VBMs transfer from the $\Gamma$ to $K$ (or between $K$ and $\Gamma$) point. 
Similarly, this direct-to-indirect transition was observed in the monolayer system~\cite{liao2020tunable}. 
Moreover, as shown in Figure~\ref{fig:u-u}g, the overall descending trends of the band gaps from compressive to tensile strains are consistent with the results of the monolayer system as well. 
However, for the D--U model, a small deviation of the linearly descending trend from 4 to 6\% strains is seen, owing to the fact that the VBM ($\Gamma$) is spatially distributed in the interlayer facing region (Figure~\ref{fig:pot}e). 
The strain-enhanced interlayer effects lead to a larger split of the CBM edges and a smaller decrease of the overall band gap. 
We further calculate the electron effective masses from the band structures, as shown in Figure~\ref{fig:u-u}h.
It is worth noting that the electron effective masses preserve isotropic with different biaxial strains for the three models, so the interlayer charge redistribution has almost no influence on the isotropy of the CBM curvature.
However, significant nonlinear deviations (comparing to the monolayer) are seen specially for the U--U and D--U models.
The reasons are not only because of the vdW interlayer effects, but also more profound changes in the ionic lattices.
We will discuss this fact in detail in the next Section.

\begin{figure*}[ht!]
 \centering
 \includegraphics[width=17.2cm]{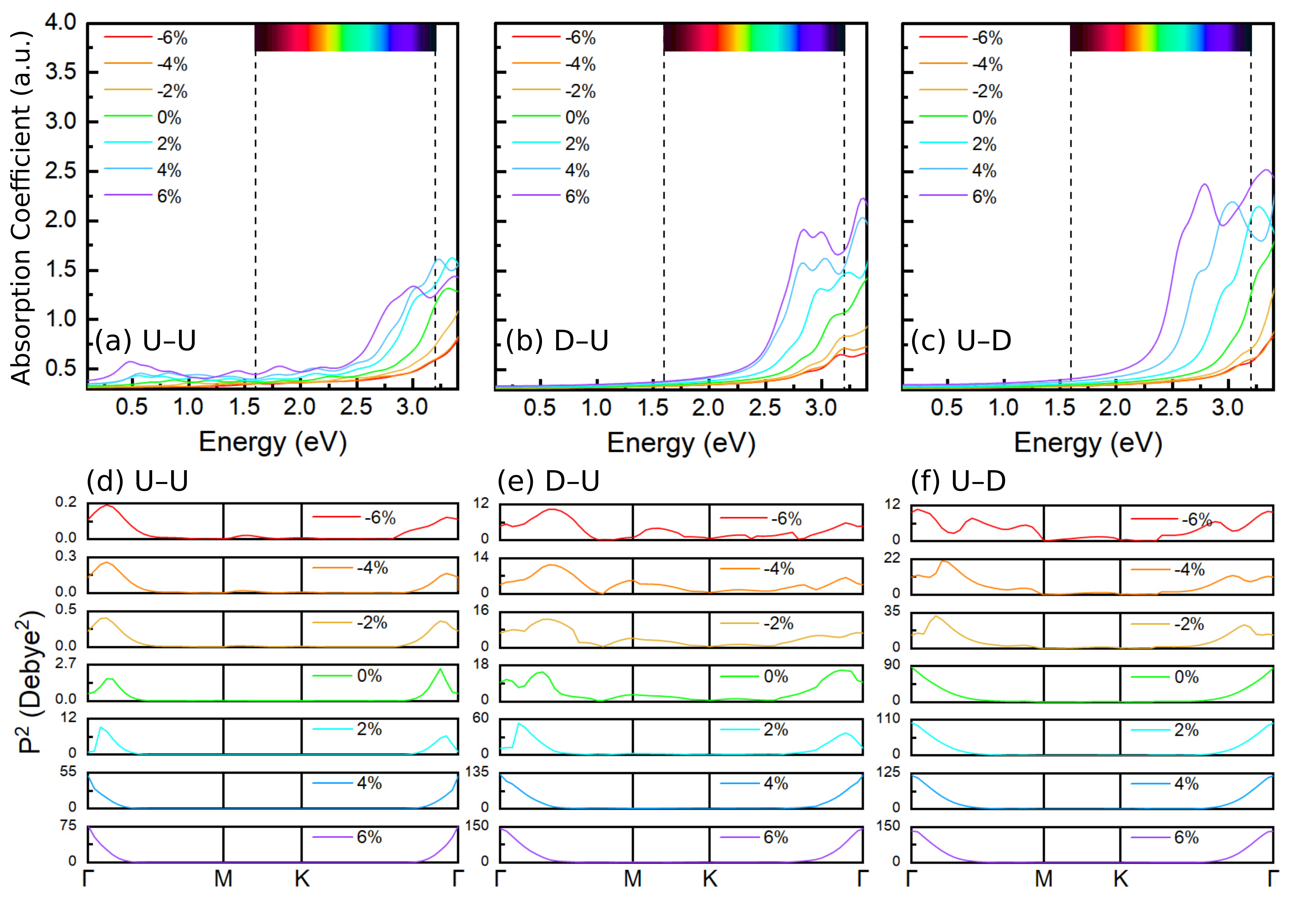}
 \caption{(a-c) Linear optical absorption spectra of the three models. Visible light range is labeled by the dashed lines from 1.63 to 3.20 eV. (d-f) The transition dipole moments of the three models from VB to CB edges. The color coding of the curves indicates the biaxial strain from -6 to 6\%.}
 \label{fig:pho}
\end{figure*}

We also probe the stain-engineered linear optical properties with the static dielectric matrix derived using the projector-augmented wave (PAW) methodology~\cite{gajdos2006linear}.
The normal-incidence approximation and the Kramers-Kronig relation are used to calculate absorption coefficient~\cite{jahoda1957fundamental, fox2012optical}.
As shown in Figure~\ref{fig:pho}a-c, the results show clear red-shifts upon the increasing tensile strains for the three models, which is in good agreement with the narrowing trends of the band gaps.
The violet/blue-light absorption is much enhanced by the tensile strains. 

Figure~\ref{fig:pho}d-f show the squared dipole transition moments~\cite{meng2017parity}, $\emph{P}^{2}$, from the VB to CB edges.
It can provide useful insights on the transition probability of photoexcited electron. 
For the three models, the strongest interacting $k$-points (the highest peaks of the $\emph{P}^{2}$) transfer from non-$\Gamma$ to $\Gamma$ point upon decreasing compressive strain and increasing tensile strain, despite of the fact that the VBMs do not locate at $\Gamma$ points. 
The optical transition probabilities from the VB edge to CB edge are determined by (a) the energy differences between the VB and CB edges, and (b) the spatial overlapping of their wavefunctions. 
Therefore, even though the U--U model has the smallest band gap among the three models, it has the overall smallest $\emph{P}^{2}$ and absorption coefficient, due to the large spatial separation of the VB and CB edges which are localized on the two monolayers (Figure~\ref{fig:pot}). 
On the other hand, the non-$\Gamma$-to-$\Gamma$ transitions of the $\emph{P}^{2}$ maxima are in good accordance with the increasingly overlapped wavefunctions of the VB and CB edges at the $\Gamma$ point.
We note that by combining the different dipole models with the strain engineering, high tunablility can be achieved for the optical properties of the bilayer systems.   

\subsection{Biaxial strain induced interlayer transition} 
\label{sec:transition}

\begin{figure*}[ht!]
 \centering
 \includegraphics[width=17.2cm]{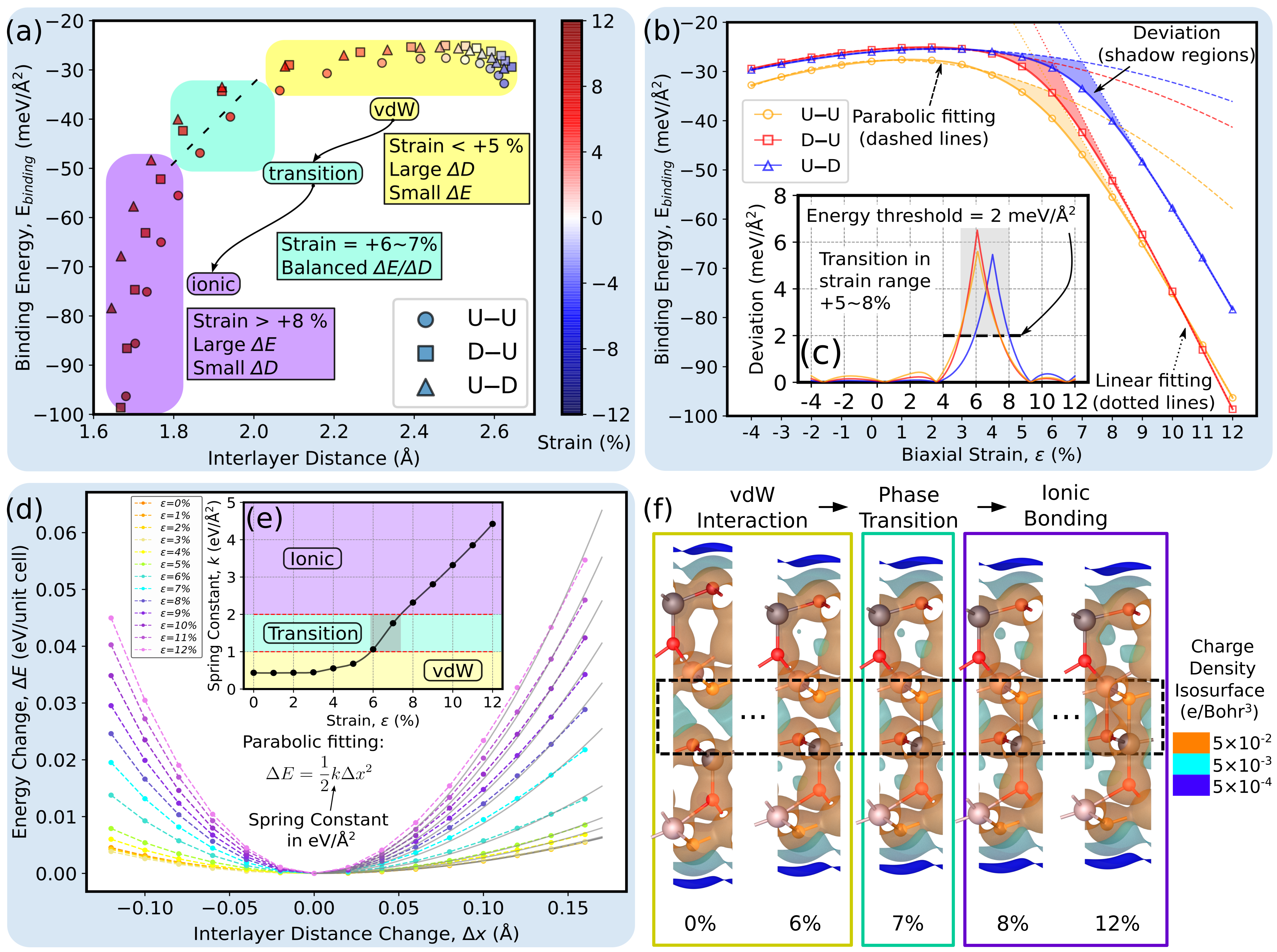}
 \caption{Analyses on the transition of the interlayer interaction: (a) the correlation between interlayer distance and binding energy; (b,c) the correlation between biaxial strain and binding energy; (d,e) the bonding force constant (bonding strength) of the interlayer interaction in the U--U model; (f) the evolution of the charge density of the U-U model.}
 \label{fig:transition}
\end{figure*}

Although bulk \ce{Ga2O3} polymorphs are ionic materials~\cite{pearton2018a, ponce2020structural, swallow2020influence}, our calculations so far show that the FE-ZB$'$ \ce{Ga2O3} bilayer systems exhibit typical vdW interlayer interactions in relaxed and moderate strained states from -6 to 6\%. 
However, with increase of biaxial tensile strain, an unexpected phase transition of the interlayer interaction appears around 7\% strain. 
As shown in Figure~\ref{fig:transition}, to shed light on the mechanism of the transition, we closely analyze the changes of interlayer distance, binding energy, spring force constant (bonding strength) and charge distribution.

First, in Figure~\ref{fig:transition}a, the plot of the binding energy against the interlayer distance reveals three distinct interaction regions: (i) the vdW region where large changes of the interlayer distances (large $\Delta D$) lead to small changes of the binding energies (small $\Delta E$); (ii) transition region, balanced $\Delta E$/$\Delta D$; and (iii) the ionic region where small $\Delta D$ results in very large $\Delta E$. The dashed line in Figure~\ref{fig:transition}a has the slope of $\sim$100 meV/\r A$^{3}$ and the transition region (cyan region) has the biaxial strains of 6$\sim$7\%. 
Similarly, the correlation between the binding energy and biaxial strain indicates a wider transition region from 5 to 8\%. 
As shown in Figure~\ref{fig:transition}b, the shadow regions stand for the deviation between the calculated $E_\mathrm{binding}$ and the parabolic/linear fitting curves. The inset figure (Figure~\ref{fig:transition}c) shows that the transition region lies between 5$\sim$8\% strains, if the threshold of the energy derivation is set to be larger than 2 meV/\r A$^{2}$.  
The binding energies after the transition are much larger than the reference value of $\sim$-20 meV/\r A$^{2}$ reported in Refs.~\citenum{bjorkman2012van, mounet2018two} for conventional 2D vdW materials. Therefore, the interlayer interaction from 8 to 12\% strains should be considered as strong bonding with high ionicity, \textit{i.e.}, ionic bonding.

As shown in Figure~\ref{fig:transition}d,e, the further calculations on the interlayer bonding force constants provide consistent evidence to this finding. 
As an example, the interlayer distances (\textit{i.e.}, the vertical distance between two interlayer $C_{\ce{O}}$ atoms) in the U--U model are changed from the relaxed states and the internal ionic positions are further relaxed with the fixed interlayer distances and biaxial strains.   
The energy change, $\Delta E$, varies as a quadratic function with harmonic approximation. 
Therefore, the bonding force constants, $k$, under different strains can be fitted with the harmonic function $\Delta E = (1/2) k \Delta x^{2}$, as shown in Figure~\ref{fig:transition}d, and the resultant force constants are shown in Figure~\ref{fig:transition}e.
The U--U model has a quite small bonding force constant of $~$0.5 eV/\r A from 0 to 5\% strains, while it increases quickly to 2.3 eV/\r A at 8\% strain. 
In fact, this value is comparable to the bond-stretching force constants in $\beta$-phase \ce{Ga2O3} which is 3.5 eV/\r A$^{2}$ for the weakest ionic bond~\cite{mu2020first}. 

Figure~\ref{fig:transition}f illustrates how the charge redistributes upon the tensile strain in the U--U model. Here we set the threshold length of the interlayer \ce{Ga}-\ce{O} bond to 2.3 \r A (not the interlayer distance), because it is the longest \ce{Ga}-\ce{O} bond length within a 12\%-strained monolayer. 
The orange-colored isosurfaces represent the charge density of 0.05 $e$/Bohr$^{3}$ increasing toward atom centers. 
It is clear that, at the relaxed state, the charge mainly distributes along the ionic bonds within the two monolayers, and the interlayer facing region has a gap with much less dense charge distribution (cyan isosurface). 
With the increase of the tensile strain, the charge starts to accumulate to the interlayer facing region as labeled by the black dashed box. 
In particular, the charge gradually centralizes at the $C_{\ce{O}}$-$A'_{\ce{Ga}}$ facing site, appearing as the orange isosurfaces merge together in this region.
At the 12\% strain, the $B_{\ce{Ga}}$-$C'_{\ce{O}}$ site has bond formation as well. The detailed charge distributions for the three models are shown in Supporting Information Figure S10-S12. 
Very similar trends can be seen in the D--U and U--D models with insignificant effect of the different dipole arrangement of the bilayer systems.
A mapping of charge density difference for the U--U model evidences the same conclusion as shown in Supporting Information Figure S13.    

\begin{figure*}[ht!]
 \centering
 \includegraphics[width=17.2cm]{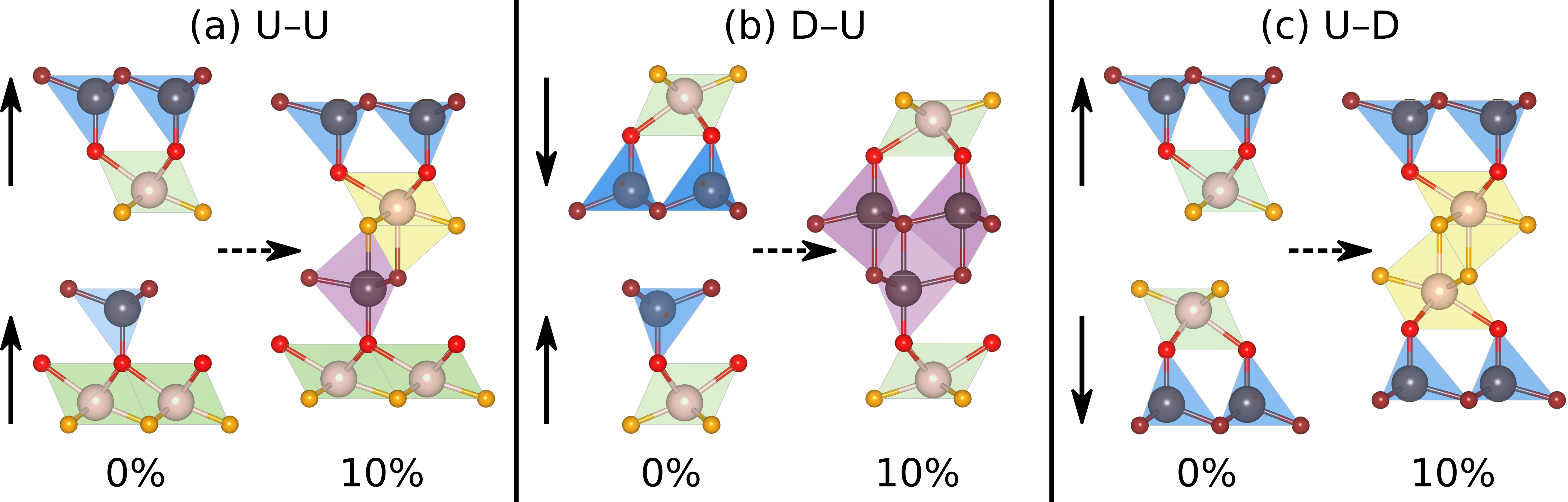}
 \caption{The transition of the folding as illustrated by the polyhedra. The colors of the polyhedra indicate the folding coordination numbers: 4-fold (blue), 5-fold (purple), 6-fold (green) and 7-fold (yellow).}
 \label{fig:fold}
\end{figure*}

In 2D vdW materials, the strain-tunable interlayer interaction and coupling is an intriguing phenomenon. 
Previous studies on other systems such as \ce{MoS2} bilayer~\cite{lee2017strain}, \ce{MoS2}/\ce{WS2} heterobilayer~\cite{pak2017strain}, and a few layers of black phosphorus~\cite{huang2019strain}, showed either slightly enhanced interaction within vdW regime or even weakened interaction upon tensile strain. 
However, here a vdW-ionic bonding transition is seen within a narrow stain range from 6 to 7\%. 
Unlike the other vdW systems, strong ionic nature of the \ce{Ga}-\ce{O} bond (Pauling's ionicity of \ce{Ga2O3}, $\sim$0.49~\cite{ma2016intrinsic}) leads to the fact that the lattice structure of \ce{Ga2O3} is governed by two principle factors: (i) the relative charges of the ions and (ii) their relative size (folding factor). 
Our previous study~\cite{SFzhao2021phase} shows that the average charge transfer from \ce{Ga} to \ce{O} atoms is about 1.80$e$ per \ce{Ga} atom for both 2D and bulk phases, 
and the big \ce{Ga} cations are energetically favorable to occupy tetrahedral (4-fold) and octahedral (6-fold) sites surrounding by the \ce{O} anions~\cite{swallow2020influence}.  
Therefore, the mechanism of the transition can be rationalized by the folding switching of the interlayer facing \ce{Ga} atoms, as shown in Figure~\ref{fig:fold}. Initially, the top and bottom \ce{Ga} atoms occupy the slightly distorted tetrahedral (blue) and octahedral (green) sites, respectively. 
These stable building blocks with high symmetry are also commonly seen in bulk \ce{Ga2O3} polymorphs, so the facing \ce{O} atoms lead to a vdW interaction between the two monolayers at the zero strain state.  
With the increase of the biaxial tensile strain (\textit{e.g.}, up to 10\%), the tetra- and octahedra inside the interlayer facing region are stretched horizontally, thus lead to a symmetry breaking of the original folding configurations. 
These asymmetric folding sites are unstable and will be compensated by the \ce{O} atoms from the other facing monolayer. 
In particular, as shown in Figure~\ref{fig:fold}, a 4-fold tetrahedron transfers to a 5-fold double tetrahedron, and a 6-fold octahedron transfers to a combined 7-fold polyhedron (an octahedron face-to-face contacting with a tetrahedron).
Depending on the different dipole models, the final combinations of the folding sites are different. This also explains the differences of the binding energies and the interlayer distances after the transition, as shown in Figure~\ref{fig:transition}a,b. It is worth noting that the high-symmetric 5-fold site is more stable than the 7-fold site, as the U--U and D--U models have larger binding energies and shorter interlayer distances after the transition. 
This strain-induced transition of the interlayer interaction provides a theoretical basis for future synthesizing and characterizing \ce{Ga2O3} atomic-thin nanolayer experimentally.

\section{Conclusion}
  
In this contribution, we systematically study the FE-ZB$'$ \ce{Ga2O3} bilayer systems using state-of-the-art \textit{ab-initio} computational methods. 
For structural properties, the stable stacking configurations and their connections via polarization reversal transition are revealed by a generalized symmetry group analysis and notations. 
The binding energies and interlayer distances are determined by the interlayer facing sites. 
For electronic properties, we first investigate the enhanced interlayer effects caused by the spatial distribution of the VBM and CBM at the relaxed state. 
Moreover, the strain-engineered band gap narrowing, red-shifts of the optical absorption spectra are investigated.
In final, we discover a unique phase transition of the interlayer interaction from vdW to ionic bonding appearing at 6$\sim$7\% tensile strain. 
The detailed analyses reveal that, the transition can be explained by a mechanism involving the switches of the folding coordinates of the interlayer facing \ce{Ga} atoms rooted from the ionic nature of the \ce{Ga2O3} systems. 
In a board perspective, the presented theoretical framework can be extended to a multitude of similar 2D vdW \ce{III2-VI3} materials. The transition mechanism can provide an important block to roadmap of experimental synthesis and application of different 2D \ce{Ga2O3} nanolayers.

\begin{acknowledgments}

This work was supported by the High-Level University Fund (G02236002 and G02236005) at the Southern University of Science and Technology. 
This computational work was supported by the Center for Computational Science and Engineering at the Southern University of Science and Technology. 
The authors are grateful to Dr. Honghao Gao at Michigan State University for the insightful discussion on symmetry group theory and notation.
\end{acknowledgments}

\bibliographystyle{apsrev4-2}
\bibliography{final}

\end{document}